\documentstyle[epsfig]{mn2e}
\def\be{\begin{equation}}
\def\ee{\end{equation}}

\begin{document}

\title{Contribution to diffuse gamma-rays in the Galactic center
region from unresolved millisecond pulsars}

\author[Wang, W., Jiang, Z.J. and Cheng, K.S.]
       {Wang, W., Jiang, Z.J. and Cheng, K.S. \\
Department of Physics, The University of Hong Kong, Pokfulam Road,
Hong Kong}

\maketitle

\begin{abstract}
The diffuse gamma-rays in the Galactic center region have been
studied. We propose that there exists a population of millisecond
pulsars (MSPs) in the Galactic Center, which will emit GeV
gamma-rays through the synchrotron-curvature radiation as
predicted by outer gap models. These GeV gamma-rays from
unresolved millisecond pulsars probably contribute to the diffuse
gamma-ray spectrum detected by EGRET which displays a break at a
few GeV. We have used the Monte Carlo method to obtain the
simulated samples of millisecond pulsars in the Galactic center
region covered by EGRET ($\sim 1.5^\circ$) according to the
different period and magnetic field distributions from the
observed millisecond pulsars in the Galactic field and globular
clusters, and superposed their synchrotron-curvature spectra to
derive the total GeV flux. Our simulated results suggest that
there probably exist about 6000 unresolved millisecond pulsars in
the region of the angular resolution for EGRET, whose emissions
could contribute significantly to the observed diffuse gamma-rays
in the Galactic center.

\end{abstract}

\begin{keywords}
Galaxy: center -- gamma-rays: theory -- pulsar: general --
radiation mechanisms: nonthermal
\end{keywords}

\pagebreak

\section{INTRODUCTION}
Gamma-ray emissions have been detected from the Galactic center
(GC) region (Churazov et al. 1994). EGRET on board the {\em
Compton GRO} has identified a central ($<1^\circ$) $\sim 30 {\rm
MeV}-10$ GeV continuum source (2EG J1746-2852) with a luminosity
of $\sim 10^{37}{\rm erg\ s^{-1}}$ (Mattox et al. 1996). Further
analysis of the EGRET data obtained the diffuse gamma ray spectrum
in the Galactic center. Allowing for a total source-excess extent
up to 1.5 degree in radius, the luminosity ($>100$ MeV) attributed
to the source excess at GC is about $2\times 10^{37}{\rm erg\
s^{-1}}$ (Mayer-Hasselwander et al. 1998). The photon spectrum can
be well represented by a broken power law with a break energy at
$\sim 2$ GeV. Below this energy the differential photon spectrum
is $F(E)=2.2\times 10^{-10}(E/1900{\rm MeV})^{-1.3}$, above the
break energy the spectrum is $F(E)=2.2\times 10^{-10}(E/1900{\rm
MeV})^{-3.1}$. A re-analysis of EGRET data by Hooper and Dingus
(2002) has shown that the EGRET GeV source is displaced from the
GC. Recently, Tsuchiya et al. (2004) have detected sub-TeV
gamma-ray emission from the direction of the Galactic Center using
the CANGAROO-II Imaging Atmospheric Cherenkov Telescope. Their
data suggest that the GeV source 3EG 1746-2851 may be coincident
with this TeV source. Recent observations of the GC with the air
Cerenkov telescope HESS (Aharonian et al. 2004) have shown a
significant source centered on Sgr A$^*$ above energies of 165 GeV
with a spectral index $\Gamma=2.21\pm 0.09$.

Some researchers have studied the possible origin of the
gamma-rays from GC. Mastichiadis \& Ozernoy (1994) provided that
the gamma-rays originate close to the massive black hole ($M_{\rm
BH} \sim 10^6 M_\odot$), possibly from relativistic particles
accelerated by a shock in the accreting plasma. In the same time,
the gamma-rays could come from some extended features like radio
arcs, where relativistic particles are present (Pohl 1997).
Markoff et al. (1997) discussed in detail the gamma-ray spectrum
of GC produced by synchrotron, inverse Compton scattering, and
mesonic decay resulting from the interaction of relativistic
protons with hydrogen accreting onto a point-like sources (e.g.
the massive black hole).

However, the above models cannot produce the hard gamma-ray
spectrum with a sharp turnover at a few GeV, which is observed for
the GC source. On the other hand, the spectrum is similar with the
gamma-ray spectrum emitted by middle-aged pulsars and millisecond
pulsars (Zhang \& Cheng 2003; Cheng et al. 2004a). Because of the
high gamma-ray luminosity observed in GC, we may require a pulsar
population in the inner region. In the following, we will first
argue that canonical pulsars (including young and mature pulsars)
may not be a major contributor to the pulsar population in GC.

Young pulsars are not likely to be a major contributor since few
supernova remnants are presently observed in the Galactic center
field targeted in the deep X-ray surveys (Wang et al. 2002a; Muno
et al. 2003). This viewpoint is also supported by pulsar birth
rate estimates. Specifically, the birth rate of young pulsars in
the Milky Way is about 1/150 yr (Arzoumanian, Chernoff, \& Cordes
2002).  As the mass in the inner 20 pc of the Galactic center is
$\sim 10^8 {\rm ~M}_{\odot}$ (Launhardt, Zylka, \& Mezger 2002),
the birth rate of young pulsars in this region is only $10^{-3}$
of that in the entire Milky Way, or $\sim$ 1/150 000 yr. We note
that the rate may be increased to as high as $\sim 1/15000$ yr in
this region if the star formation rate in the nuclear bulge was
higher than in the Galactic field over last $10^7 - 10^8$ yr (see
Pfahl et al. 2002). Few young pulsars are likely to remain in the
Galactic center region since only a fraction ($\sim 40\%$) of
young pulsars in the low velocity component of the pulsar birth
velocity distribution (Arzoumanian, Chernoff, \& Cordes 2002)
would remain within the 20 pc region of the Galactic center
studied by Muno et al. (2003) on timescales of $\sim 10^5$ yrs.
Mature pulsars can remain active as gamma-ray pulsars up to 10$^6$
yr, and have the same gamma-ray power with millisecond pulsars
(Zhang et al. 2004; Cheng et al. 2004a), but according to the
birth rate of pulsars in GC, the number of gamma-ray mature
pulsars is not higher than 10.

On the other hand, there may exist a population of old neutron
stars with low space velocities which have not escaped the
Galactic center (see Belczynski \& Taam 2004). Such neutron stars
could have been members of binary systems and been recycled to
millisecond periods, having formed from low mass X-ray binaries in
which the neutron stars accreted sufficient matter from either
white dwarf, evolved main sequence star or giant donor companions
(Belczynski \& Taam 2004, in preparation). The current population
of these millisecond pulsars may either be single (having
evaporated its companion) or have remained in a binary system.
Cheng et al. (2004b) provide that wind nebulae of the millisecond
pulsar population in the Galactic center region can contribute to
the unidentified X-ray sources in GC by the {\em Chandra} survey
(Wang et al. 2002a; Muno et al. 2003). Because millisecond pulsars
also remain active as gamma-ray pulsars, radiating gamma-rays
through the synchrotron-curvature process (Cheng \& Zhang; 1996;
Zhang \& Cheng 2003), it is possible that the observed gamma-ray
luminosity in GC may be produced through an accumulation of these
millisecond pulsars which would provide the observed gamma-ray
spectrum.

In this paper, we will examine in detail if the millisecond pulsar
population could contribute to the diffuse gamma-ray spectrum in
the Galactic center region. In \S 2, we will present our
motivation to consider the contribution of millisecond pulsars to
diffuse gamma-ray in GC based on the outer gap model. To find the
gamma-ray spectrum of these millisecond pulsars, we assume their
globular parameters like the period, magnetic field are similar to
those of the observed millisecond pulsars. In \S 3, we derive the
period, magnetic field distributions of total MSPs, MSPs in the
Galactic field and in globular clusters from the present pulsar
survey database. In \S 4, we sample millisecond pulsars by the
Monte Carlo method according to the different distributions,
superpose their spectral profiles to fit the observed diffuse
gamma-ray spectrum in GC and obtain the number of MSPs which are
needed in the Galactic center region covered by EGRET. Our results
are summarized and discussions are also presented in \S 5.

\section{MOTIVATIONS}
Since the mass within the region with the radius $\sim 20$ pc
($17'\times 17'$) of the Galactic center is $\sim 10^8 M_\odot$
(Launhardt, Zylka \& Mezger 2002), an estimate for the fraction of
millisecond pulsars in this region is about $10^{-3}$ of the
entire Galaxy. Based on the population analysis of Lyne et al.
(1998), the number of millisecond pulsars in the entire Galaxy may
exceed $2\times 10^5$, suggesting that more than 200 millisecond
pulsars exist in the Galactic center region if their evolutionary
formation channels are similar to the rest of the Galaxy. Coupled
with the fact that the escape velocity from the Galactic center is
about $200 {\rm km\ s^{-1}}$ and the average birth velocities of
observed millisecond pulsars are $\sim 130 {\rm km\ s^{-1}}$ (Lyne
et al. 1998), these pulsars are likely to remain in the Galactic
center through their entire lifetime. Belczynski \& Taam (2004)
have considered the binary population synthesis in the Galactic
center region, and their results show that there exist about 300
low-mass binary systems in the Galactic center. Furthermore, about
100 - 200 millisecond pulsars could be produced through the
recycle scenario and lie in the Galactic center region ($17'\times
17'$, Taam 2004, private communication). Recently, Pfahl \& Loeb
(2004) propose that $\sim 1000$ radio pulsars may presently orbit
Sgr A$^*$ with periods of $\leq 100$ years, in which 1-10 may be
detected by current radio telescopes. Therefore, we believe that
there should exist a millisecond pulsar population in the Galactic
center, which can contribute to the high energy emissions, e.g.
X-rays, gamma-rays which are detectable.

Millisecond pulsars can remain active as gamma-ray pulsars through
their lifetime according to outer gap models which are originally
proposed by Cheng, Ho \& Ruderman (1986). Based on the model,
Zhang \& Cheng (1997) have developed a self-consistent mechanism
to describe the high energy radiation from spin-powered pulsars.
In the model, relativistic charged particles from a thick outer
magnetospheric accelerator (outer gap) radiate through the
synchrotron-curvature radiation mechanism (Cheng \& Zhang 1996)
rather than the synchrotron and curvature mechanisms in general,
producing non-thermal photons from the primary $e^\pm$ pairs along
the curved magnetic field lines in the outer gap. The
characteristic energy of high energy photons emitted from the
outer gap is determined by the global pulsar parameters, including
the spin period $P$, the dipolar magnetic field $B$, and the
fractional size of the outer gap $f\sim 5.5P^{26/21}B_{12}^{-4/7}$
(Zhang \& Cheng 1997), which is the ratio between the mean
vertical separation of the outer gap boundaries in the plane of
the rotation axis and the magnetic axis to the light cylinder
radius. Then the characteristic synchrotron-curvature emission
energy is given by (Zhang \& Cheng 1997) \be E_\gamma \simeq
5\times 10^7 f^{3/2}B_{12}^{3/4} P^{-7/4}({r\over R_L})^{-13/8}
{\rm eV}, \ee where $B_{12}$ is the dipolar magnetic field in
units of $10^{12}$ G, $R_L=cP/2\pi$ is the light cylinder radius,
and $r$ is the distance to the neutron star. The gamma-ray
spectrum drops exponentially beyond the energy $E_\gamma$. This
self-consistent model has also been developed to describe
gamma-ray emission from millisecond pulsars (Zhang \& Cheng 2003).

Zhang \& Cheng (1998) have studied the contribution to the
Galactic diffuse gamma-rays from the unresolved spin-powered
pulsars using the outer gap model. Their results show that the
gamma-ray emission from these pulsars could contribute
significantly to the observed Galactic diffuse gamma-ray spectrum
above 1 GeV. Therefore, we believe that a large number of
millisecond pulsars which lie in the Galactic center region could
also contribute significantly to the diffuse gamma-ray spectrum
from the Galactic center. Furthermore, according to the results of
Zhang \& Cheng (2003), the gamma-ray spectral cut-off at $\sim$ a
few GeV is consistent with the observed spectral properties of
diffuse gamma-rays in the Galactic center.

To study the contribution of millisecond pulsars to the diffuse
gamma-ray radiation from the Galactic center in detail, e.g.
fitting the spectral properties and total luminosity, we firstly
need to derive the period and surface magnetic field distribution
functions of the millisecond pulsars respectively. And then we
integrate contributions from all the millisecond pulsars with
different periods and surface magnetic fields to derive the
predicted diffuse gamma-ray spectrum, which can be compared with
the observed spectrum in the Galactic center region to calculate
how many MSPs are needed. This is the aim of the present paper.

\section{DISTRIBUTION FUNCTIONS OF MILLISECOND PULSARS}
We obtain the period and surface magnetic field distribution
functions of millisecond pulsars from the observed pulsar data.
Here, millisecond pulsars are defined as the pulsar with $P<10$
ms, and $B_{\rm surf} <10^{10}$ G. So we find 86 detected
millisecond pulsars from the latest ATNF Pulsar Catalog
\footnote{http://www.atnf.csiro.au/research/pulsar/psrcat/}, in
which 41 millisecond pulsars are in the Galactic field, and 45
pulsars in globular clusters. Since the millisecond pulsars in the
Galactic field and globular clusters may have different
properties, we derive the distribution functions of total
millisecond pulsars at first, and then the distribution functions
of the millisecond pulsars in the Galactic field and globular
clusters separately. We should notice that the period derivative
measurement of millisecond pulsars in globular clusters is quite
difficult and uncertain, so many of them have not been given the
surface magnetic field, and the magnetic field varies from $10^8$
G to $10^{10}$ G. On the other hand, the measurement of the
millisecond pulsars in the Galactic field is relatively accurate
and reliable, the derived surface magnetic field is lower than
$10^9$ G.

When fitting the normalized distribution profile, we take the
Gaussian function as the following form \be f(x)=f_0+ {A\over
W\sqrt{\pi/2}}\exp[-2({x-x_c\over W})^2]. \ee To derive the period
distribution function, we define $x=P$ and $1{\rm ms}<P<10$ ms,
and for surface magnetic field distribution function, $x=\log
B_{\rm surf}$. If including the total millisecond pulsars, the
fitting parameters for the period distribution are $f_0=0.045,
A=0.60, W=1.95, x_c=3.91$; for magnetic field distribution, the
parameters are $f_0=0.015, A=0.17, W=0.57, x_c=8.49\ (7.8< \log B<
10)$. The fitting profiles of the pulsar data have been shown in
Figure 1. Just including the millisecond pulsars in the Galactic
field, the parameters for the period distribution are $f_0=0.059,
A=0.48, W=2.34, x_c=4.28$; the parameters for magnetic field
distribution are $f_0=-0.06, A=0.24, W=0.65, x_c=8.40\ (7.8< \log
B< 9)$ (Figure 2). Only including the millisecond pulsars in
globular clusters, the parameters for the period distribution are
$f_0=0.042, A=0.67, W=1.62, x_c=3.71$; the parameters for magnetic
field distribution are $f_0=0.04, A=0.10, W=0.37, x_c=8.71\ (7.8<
\log B< 10)$ (Figure 3).

\section{SPECTRAL MODELLING OF DIFFUSE GAMMA-RAYS OF THE GALACTIC
CENTER}

As discussed in Section 2, there exists the population of
millisecond pulsars in the Galactic center region. Firstly, we
assume the number of MSPs, $N_{\rm MSP}$, in GC within the angular
resolution size of EGRET $\sim 1.5^\circ$, each of them with an
emission solid angle $\Delta \Omega \sim$ 1 sr and the
$\gamma$-ray beam pointing in the direction of the Earth. We
performed calculations to sample the parameters (period and
magnetic filed) of these MSPs by the Monte Carlo method using the
above three distributions of the observed MSPs derived in Section
3. Here, we have assumed no evolution for these millisecond
pulsars, and they will remain to lie in the Galactic center
because of their low average proper motion velocity (see Lyne et
al. 1998; Arzoumanian, Chernoff, \& Cordes 2002)

Zhang \& Cheng (2003) have proposed a model to describe the X-ray
and $\gamma$-ray emission from MSPs with outer gaps. We first
calculate the the fractional size $f_m$ of the outer gap in our
simulated MSPs, if $f_m < 1$, the outer gap can exist and then the
MSP can emit high energy $\gamma$-rays, $f_m$ can be estimated by
(Zhang \& Cheng 2003) \be f_m \approx 7.0 \times 10^{-2} P
^{26/21} _{-3}(\frac{B}{10^8 \rm{G}})^{4/7} \delta r_5 ^{2/7} \ee
where $\delta r$ is the distance where the local magnetic field
equal to the dipole field, in the following calculations, we
assume $\delta r_5 = \delta r /10^5 \sim 1$. The $\gamma$-rays are
produced in the outer gap by synchrotron-curvature radiation, the
$\gamma$-ray differential flux observed on the Earth of the $i$th
MSP can be calculated by \be F_i(E_\gamma)=\frac{1}{\Delta \Omega
d ^2} \frac{d^2 N_i}{d E_\gamma dt} \ee where $d$ is the distance
of the Galactic center to us, taken as 8.0 kpc, the solid angle of
$\gamma$-ray beam $\Delta \Omega \sim 1$ sr, the spectrum
$d^2N/dE_{\gamma} dt$ can be calculated according to the equation
(57) of Zhang \& Cheng (1997). The total flux which contributes to
the $\gamma$-ray emission from the GC region can be obtained by
superposing all MSPs with outer gaps \be F(E_\gamma)=\sum_{i=1}^n
F_i(E_\gamma) \ee where $n$ is the number of simulated MSPs with
outer gaps.

During our calculations, we let the number of millisecond pulsars
$N_{\rm MSP}$ as a free parameter to fit the observed data points
using three different distributions of the period and magnetic
field separately. We find that about 6000 MSPs could significantly
contribute to the observed GeV flux in the Galactic center region.
The calculated profiles of superposed spectra of the all
millisecond pulsars with outer gaps in the Galactic center region
according to three different distributions of the period and
magnetic field are shown in Figure 4. The solid line corresponds
to the distributions derived from the total detected millisecond
pulsars; the dashed line just includes the millisecond pulsars in
the Galactic field; and the dotted line just includes the
millisecond pulsars in globular clusters. Our predicted spectra
are consistent with the observed results which have been analyzed
by Mayer-Hasselwander et al. (1998) and Hartman et al. (1999).

In Figure 4, one can find that the predicted spectra calculated
due to the distributions derived from the total observed
millisecond pulsars and those just in globular clusters fit the
observed data better. The reduced $\chi^2$ values of three curves
fitting to the eight data points are 1.51 (solid line), 3.95
(dashed) and 1.62 (dotted) respectively. Our results probably
imply that the unresolved millisecond pulsars in GC will follow
the period and magnetic field distributions similar to the forms
observed in globular clusters. However, the number of discovered
millisecond pulsars is very limited at present, so the statistics
may be quite uncertain. In addition, the predicted flux is
dependent on some pulsar global parameters, so it is too early to
make the conclusion. Meanwhile, it should be pointed that the
different profiles of predicted spectra could also be induced by
the different number of simulated millisecond pulsars with outer
gaps (satisfying the criterion $f_m<1$): about 4000 simulated MSPs
have outer gaps according to the distributions just including the
MSPs in the Galactic field; while, about 5000 MSPs can have outer
gaps for the derived distributions of total observed MSPs and
those in globular clusters. For the distributions of MSPs in the
field, even if we increase the number to $N_{\rm MSP}\sim 10000$,
the predicted spectrum fits the data not better than the solid and
dotted lines in Figure 4. After checking the simulated data, we
find that according to the distributions in the field, the number
fraction of simulated MSPs which satisfies the criterion $f_m<1$
is small, and the average $\gamma$-ray luminosity of MSPs is
relatively low.

Furthermore, the fraction size of the outer gap $f_m$ will
determine the total flux of millisecond pulsars, i.e.
$F_\gamma\propto f_m^3$. Zhang et al. (2004) considered the effect
of the magnetic inclination angle in the calculation of $f_m$, and
found that $f_m$ for the large magnetic inclination angle can
increase to 2-3 times of the original value. Millisecond pulsars
generally have a large magnetic inclination angle (Ruderman 1991),
then they can emit the higher gamma-ray flux, so the required
number of millisecond pulsars in the region with the radius of
$\sim 1.5^\circ$ to produce the diffuse gamma-rays in GC can
decrease significantly. Even considering part of millisecond
pulsars have no outer gaps when $f_m$ is larger, the simulated
number of millisecond pulsars which can fit the spectra best is
$N_{\rm MSP}\sim 1000$ for the distributions of MSPs in globular
clusters and total observed MSPs, $N_{\rm MSP}\sim 2000$ for the
distributions of MSPs in the field. Probably, these decreasing
numbers of MSPs in GC are more reasonable both for the theoretical
prediction and observations.

\section{DISCUSSIONS AND CONCLUSION}
In the present paper, we have studied the diffuse gamma-rays in
the Galactic center region detected by EGRET. We propose that
there exists the population of millisecond pulsars in the Galactic
center, which will emit gamma-rays through synchrotron-curvature
radiation predicted in the outer gap models. The gamma-ray
spectrum of millisecond pulsars shows a break at a few GeV, which
is similar to the spectrum of diffuse gamma-rays in GC. Therefore,
these unresolved millisecond pulsars could contribute to a
significant fraction of diffuse gamma-rays in GC.

According to the period and magnetic field distributions of the
observed millisecond pulsars, we sampled the global parameters
(period and magnetic field) of the millisecond pulsar population
in the Galactic center region by the Monte Carlo method. Since the
millisecond pulsars in the Galactic field and globular clusters
may have different properties, we find three classes of
distributions which include total observed MSPs, just include the
MSPs in the field and those in globular clusters, respectively.
Then we used three possible MSP samples to calculate their
gamma-ray differential spectra, and superpose the profiles to fit
the observed data. The modelled results suggest about 6000 MSPs
are needed to match the observed gamma-ray spectrum. We also find
the superposed spectra of MSPs could fit the observed spectrum
well except for the sample derived from the distributions of the
MSPs in the field (see Figure 4), which probably suggests the
unresolved millisecond pulsars in the Galactic center follow the
distributions of the period and magnetic field similar to the
forms in globular clusters. However, because the number of
millisecond pulsars in globular clusters with the period and
period derivative measurements is limited, and the number of
millisecond pulsars in GC is also unknown, we cannot conclude that
the hypothetical millisecond population in GC could resemble
millisecond pulsars in globular clusters at present. Furthermore,
if the effect of the magnetic inclination angle is considered in
the calculation of the fraction size of the outer gap $f_m$, at
least about 1000 millisecond pulsars for the different
distribution are still required in the region of the radius $\sim
1.5^\circ$ to fit the diffuse gamma-ray spectrum in the Galactic
center.

The multiwavelength observations have shown the complex structure
in the GC region (e.g. Purcell et al. 1997; Mayer-Hasselwander et
al. 1998; Wang et al. 2002a; Maeda et al. 2003), so different
scenarios for the origin of the diffuse gamma-rays have been
considered as mentioned in \S 1 (also see Mayer-Hasselwander et
al. 1998 and references therein). Recently, Fatuzzo \& Melia
(2003) have attributed the GeV emission to $\pi^0$ decay resulting
from high energy protons interacting with the ambient matter in
Sgr A East. As pointed out in Mayer-Hasselwander et al. (1998),
those models cannot easily produce the very hard spectrum with a
sharp turnover. In this paper, we suggested that the spectrum
turnover may be contributed by the millisecond pulsar population
in GC through curvature-synchrotron radiation in the
magnetosphere. The predicted spectrum of millisecond pulsars show
a cutoff above $\sim$ 3 GeV (see Figure 4), so it cannot
contribute to the sub-TeV flux recently detected by CANGAROO-II
(Tsuchiya et al. 2004) and HESS (Aharonian et al. 2004), $L_{\rm
TeV}\sim 10^{35}{\rm erg\ s^{-1}}$. These TeV photons in GC are
possibly induced by pion decay produced through $pp$ interactions
(e.g. Tsuchiya et al. 2004 and references therein), and compact
wind nebulae of millisecond pulsars through inverse Compton
scattering (e.g. Aharonian, Atoyan \& Kifune 1997; Wang et al.
2004). According to the estimation of Wang et al. (2004), the TeV
luminosity of the MSP wind nebula through inverse Compton
scattering is about $10^{31}{\rm erg\ s^{-1}}$, with $<L_{\rm
sd}>\sim 10^{34}{\rm erg\ s^{-1}}$, the average medium density
$n\sim 10^2{\rm cm^{-3}}$ and magnetic field in the Galactic
center region $B\sim 50\mu$G (Uchida \& G\"usten 1995), then the
total TeV luminosity contributed by wind nebulae of MSPs is $\sim
6\times 10^{34}{\rm erg\ s^{-1}}$. Compared with the present
observations, we think compact wind nebulae of millisecond pulsars
through inverse Compton scattering could contribute to the TeV
flux in the Galactic center, and the photon index $\Gamma\sim 2.2$
is also well within the predicted range by wind nebula models
$\Gamma\sim 2-2.5$ (e.g. Wang et al. 2004). Hence, the origin of
high energy gamma-rays (GeV - TeV energy band) is still a mystery,
requiring the further observational constraints.

Michelson et al. (1994) have derived an upper limit to $>$ 100 MeV
luminosity of the globular cluster 47 Tuc, where over 20
millisecond pulsars have been identified and an estimated total
population could be larger than 200 (Camilo et al. 2000). Using
the EGRET observed upper limit of 47 Tuc ($\sim 1.2 \times
10^{35}$ erg s$^{-1}$ together with the estimate MSP population,
the estimated gamma-ray luminosity for individual MSP is roughly
$6 \times 10^{32}$ erg s$^{-1}$, which is lower than our estimate
in this paper. However, Cheng and Taam (2003) have studied the
X-ray properties of the MSPs in 47 Tuc, in which the average X-ray
luminosity per MSP is about factor 10 lower than those typical
MSPs in the field and the spectrum is dominated by thermal
spectrum with an unusually high polar cap temperature. They
conclude that all these unusually properties can be explained by
the fact that strong, small-scale, multipole magnetic fields exist
on the surface of MSPs in the 47 Tuc. They suggest that the
typical age of MSPs in globular cluster is much older than those
in the field, therefore the surface magnetic field buried under
the crust during the accretion phase may diffuse back to the
surface in billion year time scales. They also suggest that these
complicated surface magnetic field structure can quench the outer
gap because gamma-rays emitted from the polar gap can become pairs
in those magnetic field lines connected with outer gap (Ruderman
and Cheng 1988). Therefore they suggest that MSPs in 47 Tuc are
weak gamma-ray emitters.

Here, we would like to emphasize two important points again.
First, the evolution of magnetic field of millisecond pulsars in
47 Tuc could be very special. In section 5 of Cheng and Taam
(2003), who have argued that the density of star in 47 Tuc is so
high that there could be more than one exchange collision or tidal
capture to form a close interacting binary system in 47 Tuc.
Consequently, the spin-up (or spin-down) evolution during
accretion (or post accretion) phase of a millisecond pulsar in 47
Tuc can be very much different from the evolution of other
millisecond pulsars. They argue that the results in complicated
surface magnetic field structure quench the outer magnetospheric
gaps as suggested by Ruderman and Cheng (1988). Therefore,
the low gamma-ray flux from 47 Tuc may not imply all millisecond
pulsars in globular cluster emitting weak gamma-ray flux. Secondly,
if we
allow number of millisecond pulsars in the field increase by a
factor of 50\%, then the population of millisecond pulsars in the
field can give an equally good fit as those two previous cases.
Therefore we cannot conclude that the hypothetical millisecond
population in GC should resemble either millisecond pulsars in the
field or millisecond pulsars in globular cluster from the current
data.

We would like to remark that the spectral break of the gamma-ray
spectrum depends on three pulsar parameters, i.e. period, magnetic
field and the inclination angle ($\alpha$) of pulsars. The first
two parameters can be determined in very good accuracy, however,
the last parameter is difficult to be determined accurately. Zhang
and Cheng (2003) have calculated some model dependent gamma-ray
spectra for MSPs. They have chosen various inclination angles for
different MSPs. For PSR J0437-4715 and PSR J2124-3358, they choose
larger inclination angle ($\alpha \sim 40^{\circ}$), the model
spectral breaks occur at about 3 GeV (cf. Fig. 1 of Zhang \& Cheng
2003). On the other hand, for PSR J0218+4232 and PSR B1821-24 they
choose ($\alpha \sim 55^{\circ}$), the spectral breaks occur at
about 10 GeV(cf. Fig. 2 of Zhang \& Cheng 2003). Cheng et al.
(2004a) have also studied that effect of inclination angle on the
gamma-ray spectrum, indeed, the spectrum is quite sensitive to
this parameter (cf. Fig. 1 of Cheng et al. 2004a). In their study
they find that pulsars in the galactic plane are younger and have
larger inclination angle whereas pulsars in medium and high
galactic latitude are older and have smaller inclination angles.
They have used a Monte Carlo simulation method to show that if the
distribution of pulsar inclination angle satisfies a random
distribution (Biggs 1990), then they find that the gamma-ray
spectral break of galactic pulsars are higher than those in high
latitude by a factor of 3. Although there are very few confirmed
gamma-ray millisecond pulsars, i.e. only three possible candidates
in EGRET catalog with $3\sigma$ level (Fierro 1995) and one good
candidate PSR J0218+4232 (Kuiper et al. 2000), we believe that
high energy satellites, like Integral, GLAST, can very soon
provide more information for us about the spectral behavior of
MSPs.

Finally, we are aware that the millisecond pulsar population in
the Galactic center region is still an assumption at present
because there are no MSPs discovered in the inner region. We think
the assumption would be reasonable if compared with the observed
MSPs in the Galaxy, specially many MSPs discovered in the systems
of globular clusters, e.g. 47 Tuc (Grindlay et al. 2002), and an
estimated total number over 200 MSPs (Camilo et al. 2000). In
theory, the binary population synthesis in the Galactic center
region also suggested that there are hundreds of MSPs in GC
(Belczynski \& Taam 2004; Taam 2004, private communication).
However, because the electron density in the direction of GC is
very high (Cordes \& Lazio 2002), it is difficult to detect
millisecond pulsars by the present radio telescopes.\footnote{The
Parkes telescope has performed a short-time survey covering the
region of GC, no pulsars have been reported in the initial results
(Wang, N. \& Manchester, D. 2004, private communication).} X-ray
studies of the sources in GC would probably be a feasible method
to find millisecond pulsars by Chandra and XMM-Newton. Recent deep
X-ray surveys of the Galactic center have found a large number of
unidentified faint X-ray sources (Wang et al. 2002a; Muno et al
2003). Cheng et al. (2004b) have suggested that synchrotron X-ray
nebulae around millisecond pulsars could contribute to a fraction
of these sources, and the sources with tailed features would be
the good candidates. Some bright X-ray tails probably coincident
with radio filaments have been suggested to be the pulsar wind
nebulae after the image and spectral analyses (Wang, Lu \& Lang
2002b; Lu, Wang \& Lang 2003; Sakano et al. 2003). In a word,
though still in dispute, the millisecond population in GC is
suggestive, which could contribute to faint X-ray sources and
diffuse gamma-rays in GC as studied in this paper.

We are grateful to the referee for the comments and suggestions,
and to N. Wang, R. Manchester, R. Taam, and T.C. Weekes for the
useful discussions. This work is supported by a RGC grant of the
Hong Kong Government.

\begin{figure}
\psfig{figure=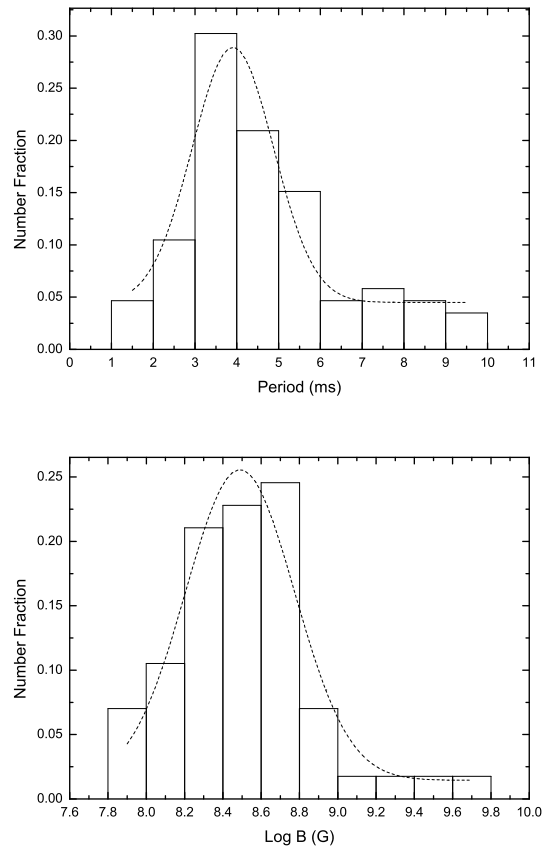,angle=0,width=10cm} \caption{The
distributions of the period (up) and surface magnetic field
(bottom) of total detected millisecond pulsars. The dashes lines
are the fitting curves using the Gaussian function.}
\end{figure}

\begin{figure}
\psfig{figure=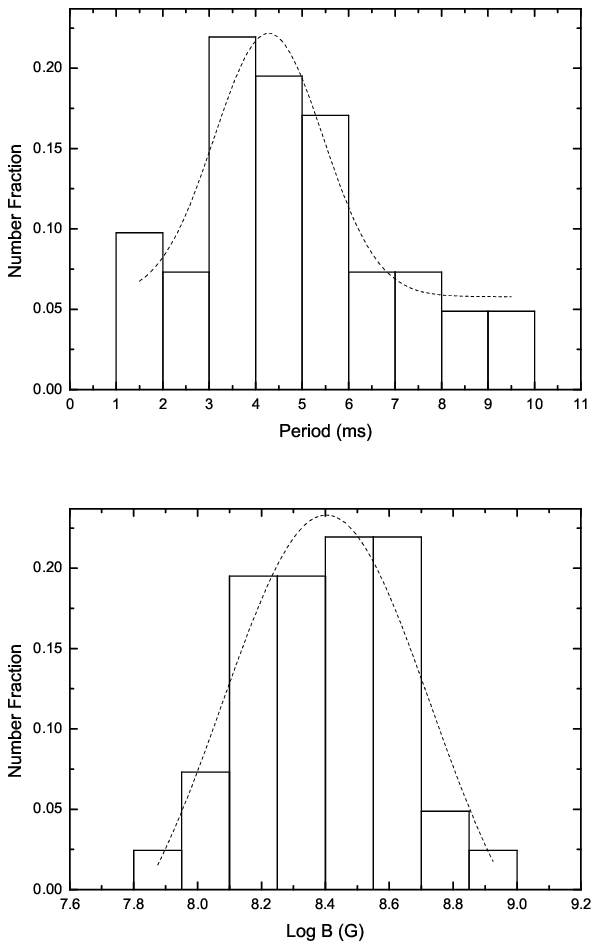,angle=0,width=10cm} \caption{The
distributions of the period (up) and surface magnetic field
(bottom) of the detected millisecond pulsars in the Galactic
field. The dashes lines are the fitting curves using the Gaussian
function.}
\end{figure}

\begin{figure}
\psfig{figure=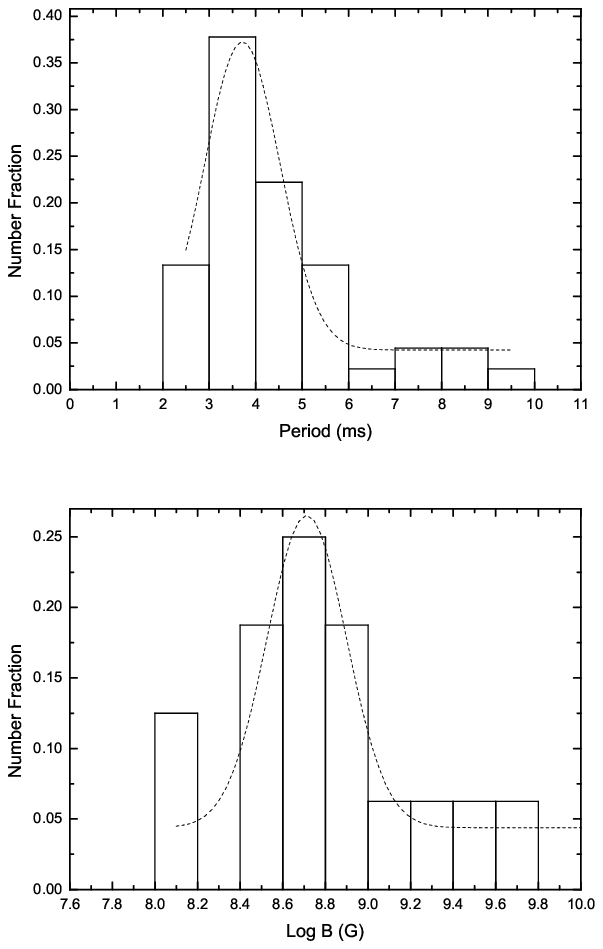,angle=0,width=10cm} \caption{The
distributions of the period (up) and surface magnetic field
(bottom) of the detected millisecond pulsars in globular clusters.
The dashes lines are the fitting curves using the Gaussian
function.}
\end{figure}

\begin{figure}
\psfig{figure=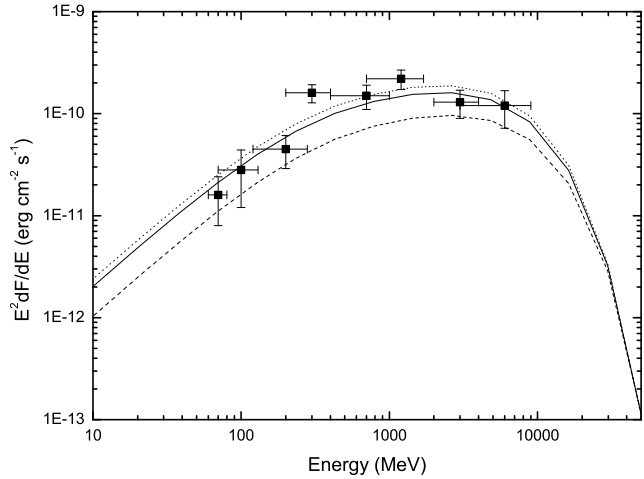,angle=0,width=10cm} \caption{Simulated
spectra of millisecond pulsars (assumed $N_{\rm MSP}=6000$) in our
model compared with observed gamma-ray energy distribution of the
Galactic center region by EGRET. The data points are taken from
the analyses of Hartman et al. (1999) and Mayer-Hasselwander et
al. (1998). The lines represent the integrated fluxes from
millisecond pulsars according to three different distributions
(period and magnetic field) of millisecond pulsars assumed in the
Galactic center, respectively. The solid line corresponds to the
distributions derived from the total detected millisecond pulsars;
the dashed line just includes the millisecond pulsars in the
Galactic field; and the dotted line just includes the millisecond
pulsars in globular clusters.}
\end{figure}

\end{document}